\documentclass{achemso}
\setkeys{acs}{doi = true}

\usepackage{amsmath,amsfonts,amssymb}
\usepackage[version=3]{mhchem}
\usepackage{graphicx}
\usepackage{setspace}
\usepackage{tocloft}
\usepackage{hyperref}
\newcommand{\doi}[1]{\href{http://dx.doi.org/#1}{\nolinkurl{#1}}}
\usepackage{xcolor}
\usepackage[margin=5mm]{caption}
\usepackage{hyperref}
\usepackage{nicefrac}
\usepackage{ulem}
\usepackage{soul}
\usepackage{svg}
\usepackage{comment}

\usepackage{achemso}
\setkeys{acs}{maxauthors = 0}

\DeclareUnicodeCharacter{03BC}{\ensuremath{\mu}}

\title{Phase behavior and electrical transport in DBTTF–HATCN donor–acceptor mixtures}

\author{Andreas Opitz}
\affiliation{Department of Physics, Humboldt-Universit\"at zu Berlin, Berlin, Germany}
\email{andreas.opitz@hu-berlin.de}
\author{Hongwon Kim}
\affiliation{Institut für Physik, Universit\"at Augsburg, Augsburg, Germany}
\author{Dmitry Lapkin}
\affiliation{Institut für Angewandte Physik, Universit\"at T\"ubingen, T\"ubingen, Germany}
\email{dmitry.lapkin@uni-tuebingen.de}
\author{Gianfranco Melis}
\affiliation{Institut für Angewandte Physik, Universit\"at T\"ubingen, T\"ubingen, Germany}
\author{Ainur Abukaev}
\affiliation{Institut für Angewandte Physik, Universit\"at T\"ubingen, T\"ubingen, Germany}
\author{Marie Siegert}
\affiliation{Experimental Physics VI, Julius-Maximilians-Universit\"at W\"urzburg, W\"urzburg,  Germany}
\author{Lennart Frohloff}
\affiliation{Department of Physics, Humboldt-Universit\"at zu Berlin, Berlin, Germany}
\author{Lisa Schraut-May}
\affiliation{Experimental Physics VI, Julius-Maximilians-Universit\"at W\"urzburg,  W\"urzburg,  Germany}
\author{Oleg Konovalov}
\affiliation{European Synchrotron Radiation Facility (ESRF), Grenoble, France}
\author{Alexander Hinderhofer}
\affiliation{Institut für Angewandte Physik, Universit\"at T\"ubingen,  T\"ubingen, Germany}
\author{Frank Schreiber}
\affiliation{Institut für Angewandte Physik, Universit\"at T\"ubingen,  T\"ubingen, Germany}
\author{Jens Pflaum}
\affiliation{Experimental Physics VI, Julius-Maximilians-Universit\"at W\"urzburg,  W\"urzburg,  Germany}
\author{Wolfgang Brütting}
\affiliation{Institut für Physik, Universit\"at Augsburg, Augsburg, Germany}
\email{wolfgang.bruetting@physik.uni-augsburg.de}


\cftpagenumbersoff{figure}
\cftpagenumbersoff{table} 
\begin{document} 

\renewcommand{\uline}[1]{\textit{#1}}
\mciteErrorOnUnknownfalse
\maketitle

\begin{abstract}

The formation of donor-acceptor complexes (DACs) between the electron donor Dibenzotetrathiafulvalene (DBTTF) and the acceptor Hexaaza\-triphenylene\-hexacarbo\-nitrile (HATCN) results in a new phase with a distinctly different crystal structure as well as new optical absorption bands below the energy gaps of the two pristine materials. X-ray scattering and atomic force microscopy provide detailed insights into the film structure and morphology by systematic variation of the mixing ratio from pristine DBTTF to pristine HATCN. The measured electrical conductivity of thin films depends in a highly non-monotonic manner on the composition of the mixture and shows significantly improved charge transport compared to the pristine films.  The temperature-dependent conductivity, charge carrier concentration, and mobility were investigated across these compositions. Surprisingly, all compositions exhibited n-type behavior, except for pristine DBTTF. This behavior is explained by the electronic structure of the mixtures, as revealed by ultraviolet photoelectron spectroscopy, which indicates that charge injection and transport occur via the lowest unoccupied molecular orbital of the DAC and HATCN. Additionally, the observed electrical conductivity is strongly influenced by morphology and structural ordering of the films. These findings offer valuable insights for the design of advanced materials with enhanced electrical performance.
\end{abstract}

\begin{spacing}{1.5}   

\section{Introduction}
\label{Introduction}  

The prospect of flexible and low-cost electronic devices made from organic semiconductors has attracted huge research interest. However, their application is limited due to, inter alia, typically lower charge carrier density and charge carrier mobility. in comparison with inorganic semiconductors. The concept of charge-transfer doping has been developed to improve electrical conductivity by using strong electron donors and acceptors as dopants for organic semiconducting materials.\cite{Baessler93,Benesi49,Nezakati18} Strong dopants typically cause a complete (or at least partial) transfer of electronic charge from one molecular species to the other, with its propensity depending on the energy landscape as well as the structural arrangement.\cite{Zhu15,Bredas04} To facilitate the transfer of charge between donor and acceptor, the ionization energy of the donor should be less than (or at least close to) the electron affinity of the acceptor, leading to ionization of both species, which is commonly termed integer charge transfer (ICT).\cite{Schwarze19} On the other hand, when the distance between donor and acceptor molecules is sufficiently close enough to allow for overlap of their $\pi$-orbitals due to a strong van der Waals interaction, orbital hybridization can occur, forming an electron donor-acceptor complex (DAC) often without noticeable charge transfer.\cite{Kitaigorodsky84} In both cases, the strong interaction between donor and acceptor molecules leads to new optical absorption features below the energy gaps of the pristine materials.

While the case of ICT has been extensively studied\cite{Schwarze19}, doping by DAC formation and its effect on electrical conduction is by far less investigated\cite{Henry15,Opitz22}, although it is the key to improving the performance of organic semiconductor devices. It can not only lead to an improvement in electrical conductivity, but can also form new crystalline phases with a lower band gap and may be intermixed with the excess donor or acceptor molecules leading to a complex phase behavior of such blends.\cite{Benesi49,Briegleb60,Mulliken62} To achieve high electric conductivity, a high doping efficiency is required, and the produced electron-hole charge carrier pairs need to overcome their mutual Coulomb attraction to become mobile. A suited approach relies on using strong dopants to generate ICT states with high charge carrier density,\cite{Tietze18} even though the doping efficiency in organic semiconductors has a limitation due to clustering or crystallite formation, resulting in a high thermal activation energy.\cite{Naab14,Henry15} 

Furthermore, a large activation energy has been reported at low doping concentration due to the presence of deep traps,\cite{Tietze15,Olthof12} and deep Coulomb potentials caused by the interaction of mobile carriers with the ionized dopants.\cite{Arkhipov05} On the other hand, as the doping concentration increases, $\pi$-orbitals begin to overlap, weakening the Coulomb force and reducing thermal activation energy.\cite{Arkhipov05_2,Mityashin12,Coropceanu07,Menke12} In addition, it has recently been shown that dielectric screening becomes more effective, since the dielectric constant of doped organic semiconductors increases with doping concentration.\cite{Koch&Blom,Koster}

\begin{figure}[tb]
\begin{center}
\begin{tabular}{c}
\includegraphics[width=8cm]{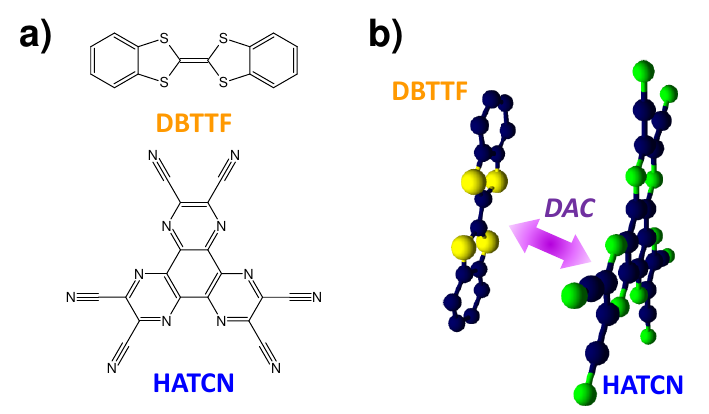}
\\
\end{tabular}
\end{center}
\caption 
{\label{fig:Molecule}
a) Chemical structures of DBTTF (electron donor) and HATCN (electron acceptor). b) Possible formation of a donor-acceptor complex (DAC) by mutually overlapping $\pi$-orbitals in a face-to-face arrangement.} 
\end{figure}

In this paper, we combine the strong electron donor DBTTF, Dibenzotetrathiafulvalene, \cite{beye+19cm, Nielsen1978} with the strong acceptor HATCN, Hexaazatriphenylenehexacarbonitrile. \cite{Lee2012c} Both are rigid planar molecules which usually supports DAC formation, however, with significant differences in molecular shape and size.\cite{Emge82,Szalay01} Figure~\ref{fig:Molecule}a shows the chemical structures of both materials.  Here, we fabricate films by co-evaporation of both species with varying mixing ratios and study the formation of DACs between them (see Figure~\ref{fig:Molecule}b), as well as their effect on electronic properties and electrical transport. For this we apply a newly developed high-throughput thin film preparation technique to perform a seamless composition gradient between both materials \cite{lapkin25} and study their structural, morphological and optical characteristics to identify signatures of new emerging DAC phase. Thereafter, we fabricate a representative set of samples with different compositions on device-relevant substrates and study their electronic structure and electrical transport properties. Depending on the mixing ratio, we find a wide range of different surface morphologies of co-deposited thin films. Additionally, the occurrence of DACs changes the electronic characteristics and, specifically, reduces the activation energy for charge carrier generation, thus increasing the density of mobile charges and, with that, the electrical conductivity as well.

\section{Results}
\label{Result}

\subsection{A: Growth studies}

To study the structure, morphology and the optical properties of the mixed DBTTF:HATCN films, we deposited a film with a compositional gradient along the substrate surface using a specially designed high-vacuum organic molecular beam deposition chamber~\cite{lapkin25}. The film with the nominal thickness of 20~nm was deposited on a silicon substrate (10~$\times$~52~mm$^{2}$) with a native silicon oxide layer and had the compositional gradient along the long substrate direction. Probing different points on the sample surface with spatially-resolved techniques, such as X-ray reflectometry (XRR), grazing-incidence wide-angle X-ray scattering (GIWAXS), atomic force microscopy (AFM) and UV-Vis spectroscopy, allowed us to directly correlate the observed properties with the composition at the given position on the substrate . The details of the measurements are given in the Methods section.

\subsubsection{Crystalline structure}

\begin{figure}[b!]
\begin{center}
\begin{tabular}{c}
\includegraphics[width=0.9\textwidth]{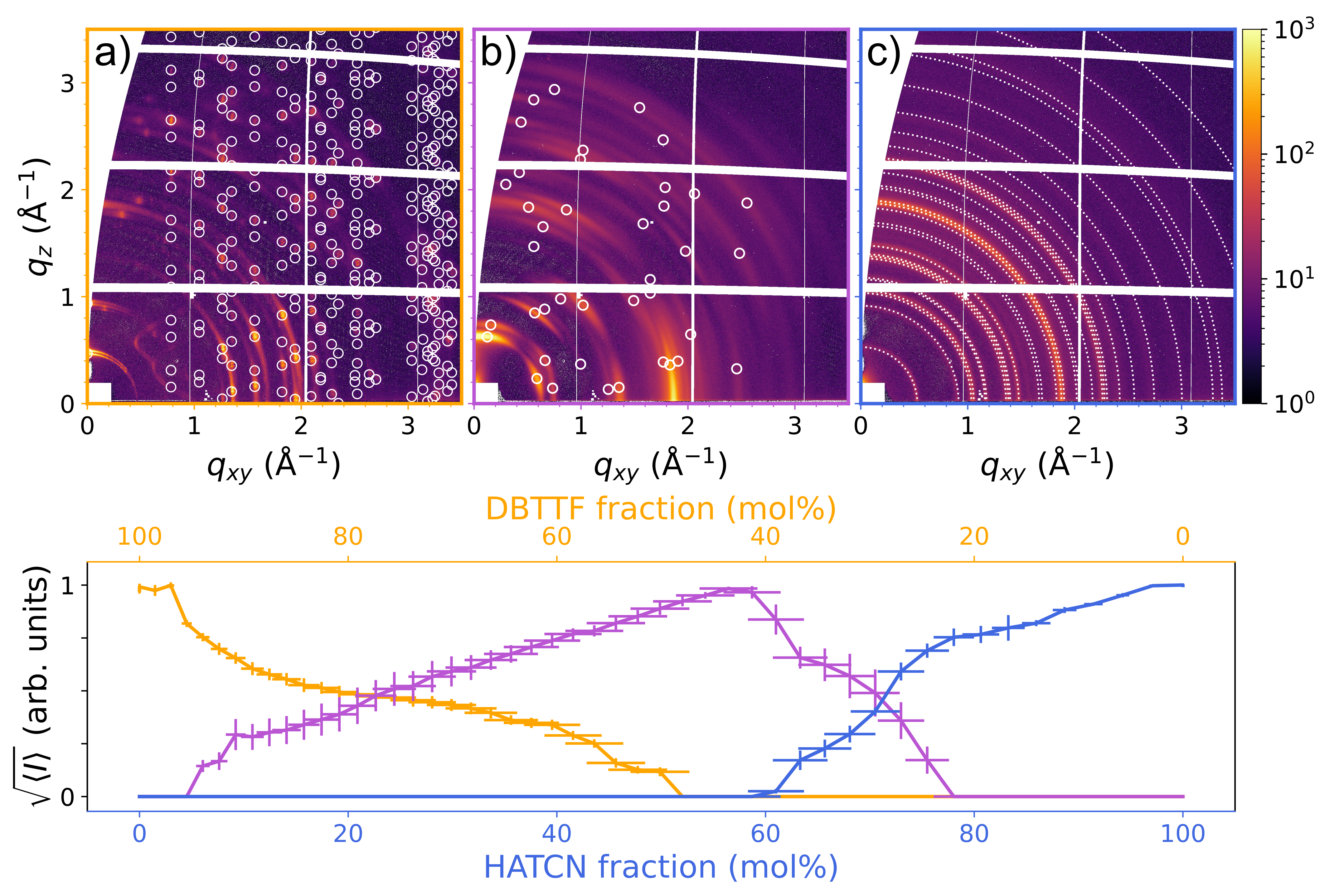}
\\
\end{tabular}
\end{center}
\caption 
{ \label{fig:GIWAXS}
Crystalline structure of the DBTTF:HATCN film with gradient in relative concentrations. (a)-(c) Reciprocal space maps measured at the DBTTF edge (a), HATCN edge (c), and in the mixed part of the sample (b). Arcs and circles indicate the simulated peak positions for the previously reported structures of the pristine DBTTF and HATCN phases and the fitted peak positions for the mixed phase. (d) Square root of the average fitted intensity of selected Bragg peaks corresponding to the crystalline phases shown in panels (a)-(c). The intensity corresponding to the pristine DBTTF phase is shown in orange, the pristine HATCN in blue, and the mixed phase in purple.} 
\end{figure}

The evolution of the crystalline structure of the mixed gradient film as a function of the composition has been investigated with GIWAXS. GIWAXS measurements were performed at the ID10-SURF beamline at the ESRF-EBS synchrotron (Grenoble, France). Measurements were taken every 1~mm along the compositional gradient of the film. Figure~\ref{fig:GIWAXS}a-c shows the reciprocal space maps corresponding to the pristine DBTTF edge, pristine HATCN edge, and the 1:1 mixed phase. 

The reciprocal space map for the DBTTF-rich edge (Figure~\ref{fig:GIWAXS}a) shows sharp Bragg peaks, indicative of a well-oriented crystalline structure. The Bragg peak positions correspond to the previously reported crystalline structure of DBTTF in thin films~\cite{Opitz22} with a triclinic unit cell containing two DBTTF molecules with the unit cell parameters of $a = 6.02$~\AA, $b = 8.05$~\AA, $c = 13.90$~\AA, $\alpha = 100.291^{\circ}$,  $\beta = 99.901^{\circ}$ and $\gamma = 94.001^{\circ}$ and the [001] plane parallel to the substrate. The peaks are split in the vertical direction, indicating that DBTTF does not form a continuous film.

In contrast, at the HATCN-rich edge (Figure~\ref{fig:GIWAXS}c), the diffraction pattern contains several Debye–Scherrer rings. The $q$-positions of the observed Debye-Scherrer rings correspond to a large trigonal unit cell containing 18 HATCN molecules with the parameters $a = b = 23.8$~\AA, $c = 15.1$~\AA, $\alpha = \gamma = 90^{\circ}$ and $\beta = 120^{\circ}$.  \cite{yadav2020} The intensity distribution of the Debye-Scherrer rings in the azimuthal direction is almost uniform, but has some modulations. These modulations can be attributed to a very weak preferred orientation of the [001] plane parallel to the substrate.

The mixed film exhibits new Bragg peaks being absent in the patterns of the pristine materials. At approximately 40–50\,mol\% of DBTTF content, the diffraction pattern contains only the new Bragg peaks, while the peaks associated with the pristine components completely disappear (Figure~\ref{fig:GIWAXS}b). This suggests the formation of a new mixed DBTTF:HATCN crystalline phase. 
The peaks are much broader than those of the pristine DBTTF phase in both directions, indicating a smaller crystallite size and a lower degree of orientation.
So far, we have not been able to resolve the crystal structure of the new phase; however, it is different from known published structures of DBTTF:HATCN cocrystals grown by vapor phase transport or from solution.\cite{valencia2025_arxiv} However, since the new structure is dominant at 1:1 mixing ratios, we assume that this molecular ratio is also present in the unit cell.

The diffraction patterns at other positions and other mixing ratios can be described as a linear combination of these three characteristic patterns. We quantified the evolution of the crystalline structure by fitting the 5 to 10 most intense Bragg peaks of each observed crystalline phase in every measured pattern. The average intensity of the Bragg peaks is approximately proportional to the volume squared of the corresponding crystalline phase. \cite{alsnielsen2011} The square root of the averaged Bragg peak intensity, shown in Figure~\ref{fig:GIWAXS}d, is therefore approximately proportional to the volume fraction of the three crystalline phases described above.

Starting from pristine DBTTF, the intensity of the DBTTF Bragg peaks remains almost constant up to approx.~5\,mol\% of HATCN content. At this concentration, the intensity of the DBTTF Bragg peaks starts to decrease significantly, while the peaks corresponding to the mixed DBTTF:HATCN phase appear. The volume fraction of the mixed phase increases almost linearly within the range of 10--55\,mol\% HATCN content. The maximum intensity of the corresponding Bragg peaks at approx.~55\,mol\% coincides with the disappearance of the DBTTF-related peaks. Within the narrow range of 50-60\% HATCN content, only peaks characteristic of the mixed phase are present. This indicates 100\,mol\% conversion into the cocrystalline structure in this compositional range. At 60\,mol\% of HATCN content, the Debye-Scherrer rings originating from the HATCN phase appear. As the HATCN content increases further, the mixed phase quickly disappears at around 80\,mol\%. Interestingly, the HATCN-related rings continue to increase in intensity within the 80--95\,mol\% HATCN content range.
We suggest that a small amount of smaller DBTTF molecules in this region can admix to the crystals with larger HATCN molecules, decreasing their crystalline order without substantially changing the structure completely.

\subsubsection{Film morphology}

\begin{figure}[htb]
\begin{center}
\begin{tabular}{c}
\includegraphics[width=\textwidth]{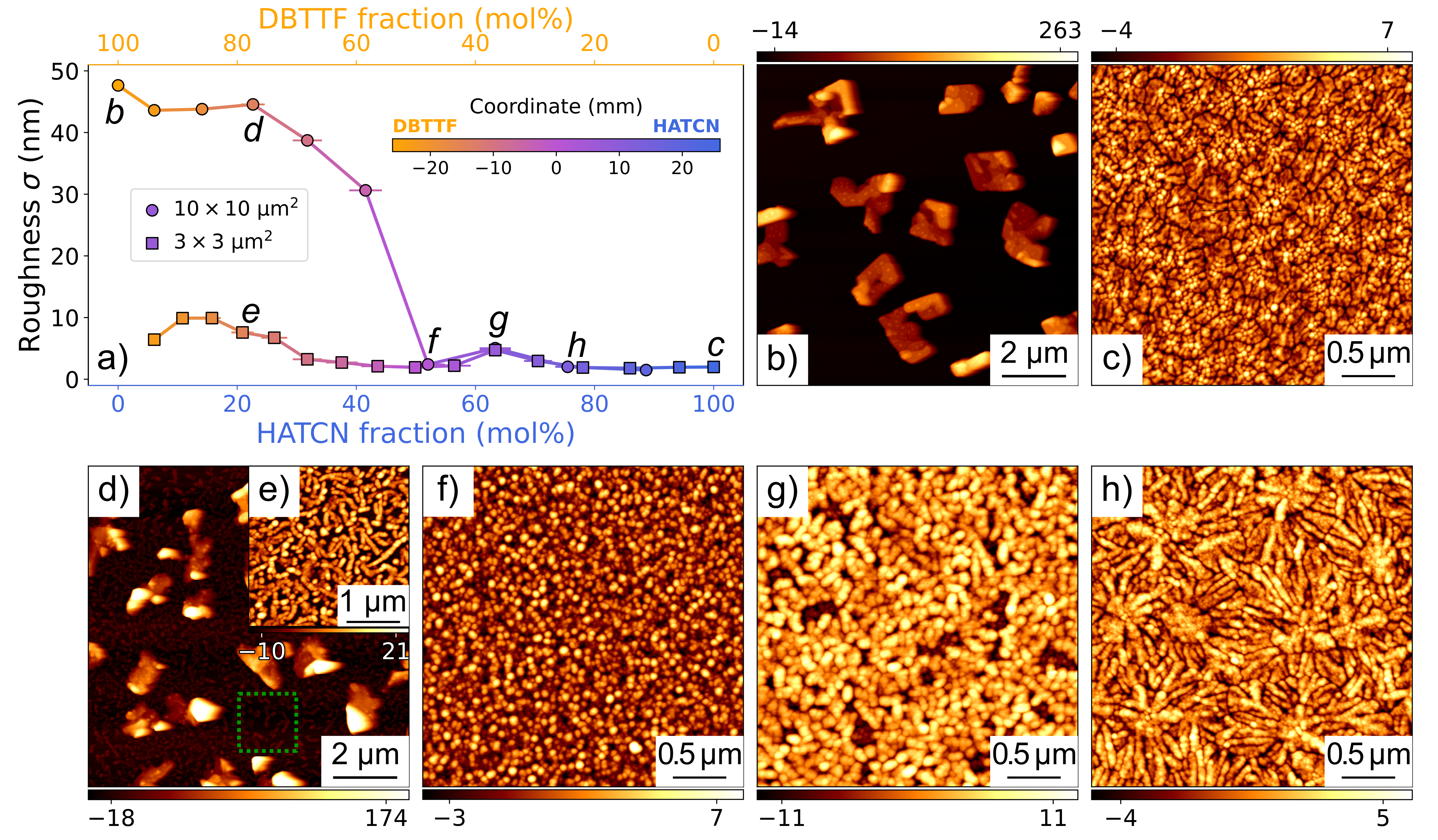}
\\
\end{tabular}
\end{center}
\caption 
{ \label{fig:AFM}
Morphology of the DBTTF:HATCN gradient film. (a) Roughness values extracted from AFM scans at different spatial positions along the gradient. The upper curve (circles) is extracted from large (10$\times$10~$\mathrm{\mu}$m$^2$) scans, the bottom (squares) one from smaller (3$\times$3~$\mathrm{\mu}$m$^2$) scans between large particles. The letters next to the curves denote the positions where the AFM maps in (b)--(h) were measured. (b),~(c) AFM maps of pristine DBTTF (b) and HATCN (c) measured at the edges of the gradient sample. (d)--(h) AFM maps measured at different positions in the mixed part of the film. Inset (e) shows a smaller scan measured between the large particles as shown in (d) with a green square. The zero value corresponds to the mean height value for each map.}
\end{figure}

Generally, the roughness and morphology evolution in two-component systems exhibits rather rich behavior. \cite{Hinderhofer2022} The morphological change in real space across the DBTTF:HATCN thin film is shown in Figure~\ref{fig:AFM}. Figure~\ref{fig:AFM}a shows the root-mean-square (RMS) roughness extracted from AFM images acquired at different positions along the compositional gradient. The dependence has two regimes in the DBTTF-rich region due to the coexistence of small and large DBTTF agglomerates, as will be discussed below. The upper region shows the values extracted from large AFM maps including the large prominent islands, the lower - from smaller maps measured between the large particles. Additional measurements of the roughness evolution with composition by means of XRR are given in the Supplementary Material and support the results shown here.

Figure~\ref{fig:AFM}b shows an AFM map measured in the pristine DBTTF region of the gradient film. DBTTF does not form a continuous film; instead, we observe large regularly shaped particles approximately 100~nm in height and 1 $\mu$m in width, typical of pristine DBTTF films grown on native silicon oxide~\cite{Yamada08}.
In contrast, at the HATCN-rich edge shown in Figure~\ref{fig:AFM}c,  we observe a continuous film with an RMS roughness of 1.5~nm, where irregular fine grains with the characteristic lateral size of 330~nm are observed, typical of pristine HATCN films~\cite{Saragi2012}.

The mixed region of the film shows a wide range of different morphological features.
As the HATCN content increases, the large crystallites characteristic of pristine DBTTF become more irregular in shape (Figure~\ref{fig:AFM}d), but remain present. At the same time, within the intergranular regions between the large islands, we observe smaller elongated crystallites with characteristic lengths of $\sim$350~nm (Figure~\ref{fig:AFM}e). 
We speculate the small crystallites in this region consist of the mixed DBTTF:HATCN phase, while the large crystallites contain mostly the excess DBTTF phase. This speculation is further supported by Raman mapping performed in this sample region, where the large crystallites are proven to contain more DBTTF (see Supplementary Information). The coexistence of large DBTTF-rich and small cocrystallites is characteristic of the entire DBTTF-rich part of the sample up to a molar fraction of about 0.5. In this region, spherical grains with a diameter of $\sim$50~nm form a continuous film with the lowest roughness of $\sim$1~nm. With increasing HATCN content, the roughness reaches a local maximum of approximately 5~nm at 0.6--0.7 HATCN molar fraction (Figure~\ref{fig:AFM}g), where the morphology changes towards the formation of larger spherical crystallites with the characteristic size of $\sim$80~nm. With further increase in HATCN content, elongated spherulite-like grain structures emerge (Figure~\ref{fig:AFM}c). The typical length of the elongated particles here reaches a diameter of about 300~nm, while the roughness decreases down to $\sim$1~nm.

\begin{figure}[b!]
\begin{center}
\begin{tabular}{c}
\includegraphics[width=\textwidth]{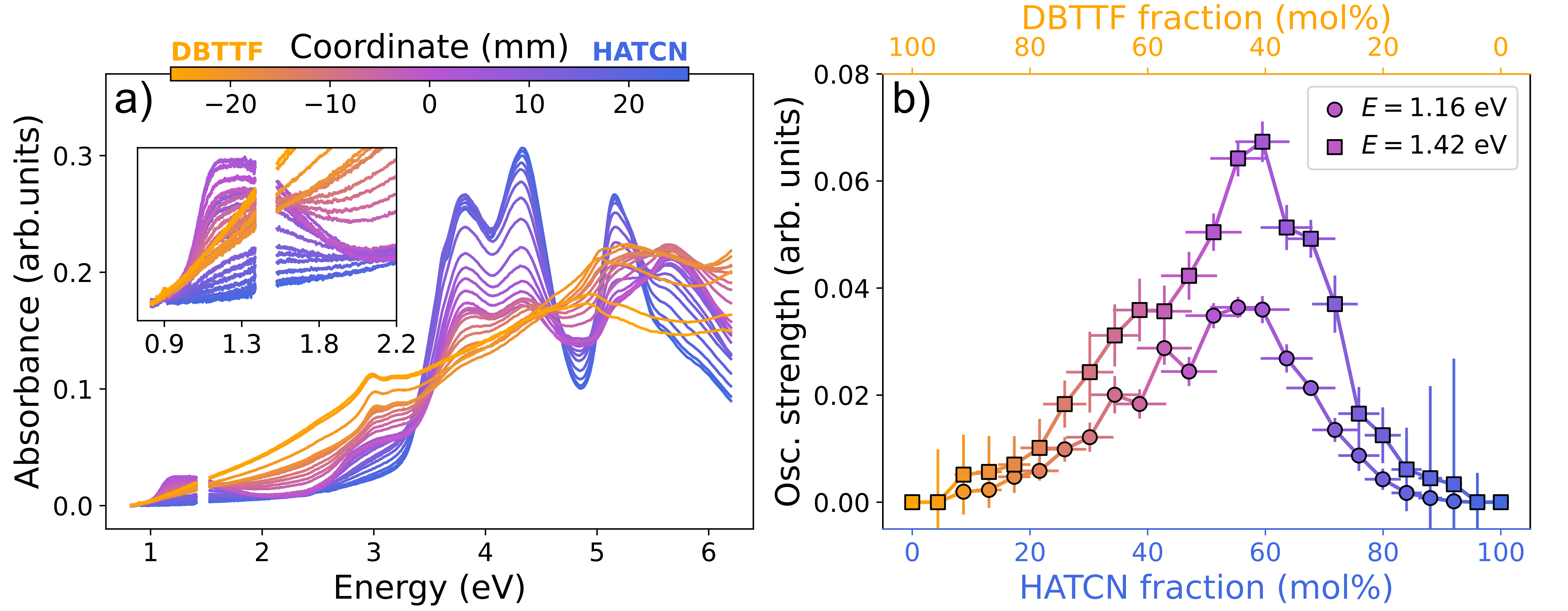	}
\\
\end{tabular}
\end{center}
\caption 
{ \label{fig:UV-Vis} (a) UV-Vis absorption spectra of the DBTTF:HATCN gradient film. The inset highlights the low-energy region of the spectra, where the CT absorption peak is located. The curves are color-coded according to the position along the film, as indicated by the color bar. (b) Extracted oscillator strength for the two CT absorption peaks at 1.16 eV (circles) and 1.42 eV (squares). } 
\end{figure}

\subsubsection{Optical properties}

The optical absorption in the ultraviolet, visible and near-infrared (UV-vis-NIR) spectral range of the DBTTF:HATCN gradient film measured at 2 mm intervals (corresponding to steps of $\sim$4\,mol\% in composition) along the compositional gradient is shown in Figure~\ref{fig:UV-Vis}a. The spectrum corresponding to the DBTTF edge is dominated by scattering rather than absorption since DBTTF does not form a continuous film, but larger isolated islands, as shown by AFM. However, two weak, distinguishable absorption peaks at ca. 3.0~eV and ca. 5~eV can be recognized. The peak positions and the spectral profile are in agreement with previous studies \cite{casado2007tetrathiafulvalene, opitz2020ordered}. At the HATCN edge, distinct absorption peaks are observed at 3.81~eV, 4.33~eV, and 5.16~eV. The peak at 3.81 eV corresponds to the lowest-energy optical transition of HATCN in accordance with previous studies \cite{johnson2016multitechnique}. 

The inset in Figure~\ref{fig:UV-Vis}a shows a magnified view of the low-energy region (0.8–2.2 eV) of the spectra, where a broad absorption peak emerges in the middle of the composition gradient. This broad peak is indicative of the DAC formation between the two molecular components. The peak was fitted with two Tauc-Lorentz oscillators with transition energies of (1.16$\pm$0.01)~eV and (1.42$\pm$0.04)~eV. The emergence of fine structure in this region may be associated with a Davydov splitting. \cite{Broch2018}
The peak positions do not change substantially with variations in composition. 
The extracted oscillator strengths for these two oscillators are shown in Figure~\ref{fig:UV-Vis}b,  as a function of the molar fraction of the two materials. A maximum is reached for both at a DBTTF:HATCN ratio of approximately 1:1, corresponding to the region of the strongest CT absorption. We note that another optical transition emerges at approximately 3~eV in the mixed region of the film and demonstrates the same behavior.

\subsection{B: Device studies}

For device studies, we grew a series of new samples on substrates with electrode structures for different types of measurements. These were ITO-covered glass substrates for photoelectron spectroscopy as well as glass substrates with patterned ITO finger electrodes for electrical conductivity measurements. For field-effect transistor structures, we used Si wafers with thermal oxide and an additional poly(methyl methacrylate) passivation layer.
Most importantly, these samples were grown at higher deposition rates of 0.4 $\sim$ 0.6 \AA/s, but without air exposure between growth and measurements. In the Supporting Information (see Figure~S4) we show exemplary AFM images of such samples to verify that film morphologies were comparable to the gradient samples used for growth studies. We want to note that due to the large difference in growth rate, some minor differences occur in the individual/respective film morphologies, but the overall behavior is qualitatively similar.

\subsubsection{Electronic structure}

\begin{figure}[htb]
\begin{center}
\begin{tabular}{c}
\includegraphics[width=\textwidth]{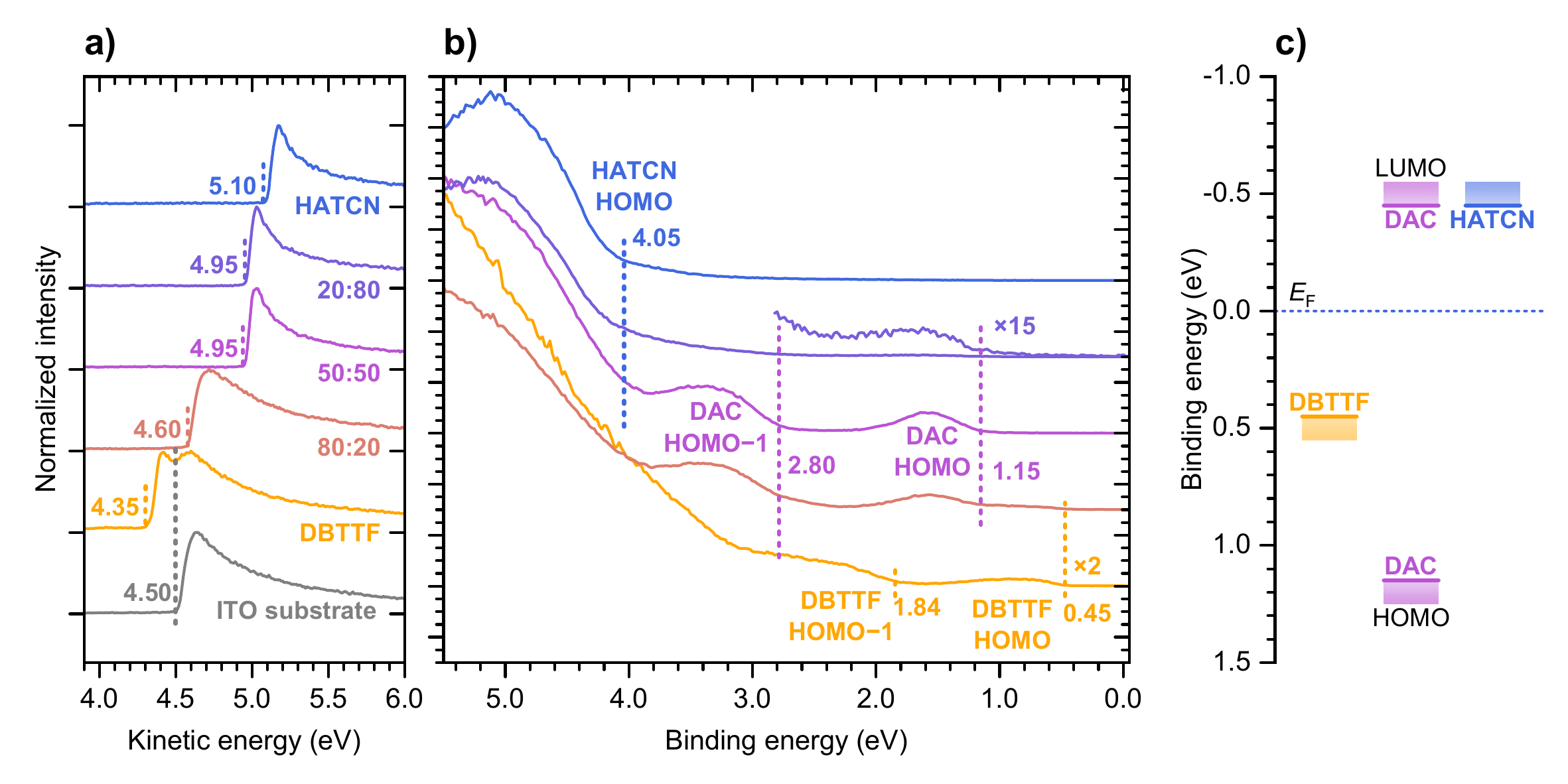}
\\
\end{tabular}
\end{center}
\caption 
{ \label{fig:UPS_Ana}
 (a) Secondary electron-cutoff (SECO) and (b) valence spectra  measured for the ITO substrate (only SECO), the pristine films and 3 mixed films. (c) Energy level diagram of frontier orbitals deduced from SECO and valence spectra (for details see text). The mixing fractions are given in mol\%.\\
  } 
\end{figure}

The valence energy levels (i.e. the highest occupied molecular orbitals, HOMO) and secondary electron cutoff (SECO) of pristine and mixed thin films (DBTTF-rich -- 4:1, equimolar -- 1:1, HATCN-rich -- 1:4) were measured by ultraviolet photoelectron spectroscopy (UPS) to understand the electronic structure and the energy level alignment on ITO electrodes. The samples were fabricated by depositing 25 nm of organic material on ITO-coated glass substrates. The obtained data is shown in Figure~\ref{fig:UPS_Ana}a. The work function of the clean ITO substrate is about 4.5 eV as determined from the onset of its SECO. Depositing pristine DBTTF on ITO results in a double SECO, that underlines the incomplete coverage as discussed in the film morphology section. The lower SECO onset represents the DBTTF work function of 4.35 eV, and the second feature is related to the work function of the ITO substrate, which is observable due to the pronounced island growth of the DBTTF film. The reduction of work function in the DBTTF region is ascribed to HOMO pinning of the DBTTF at the Fermi level of the ITO, which is induced by charge transfer between the organic material and the electrode. \cite{Oehzelt2014} In contrast, the SECO shift to higher kinetic energies after the deposition of pristine HATCN on ITO is induced by pinning at the HATCN LUMO level. \cite{Zhai20,Lee2012c} The resulting work function in this case is 5.10 eV. The work functions of the mixed films are found between these extreme cases of the pristine materials. The equimolar film and the HATCN-rich film exhibit almost the same value of the work function of 4.95 eV, and the work function of the DBTTF-rich film (4.60 ev) is closer to that of the pristine DBTTF.

Figure~\ref{fig:UPS_Ana}b shows the photoelectron spectrum of the occupied valence states for pristine and mixed films. The HATCN film shows the HOMO onset energy of 4.05 eV, which agrees with the LUMO pinning concluded from the SECO measurements.\cite{Christodoulou2014} The DBTTF film exhibits clearly separated HOMO and HOMO$-1$ states. The HOMO of DBTTF is pinned at the Fermi level as concluded before from SECO reduction. As the surface is inhomogeneously covered with molecules, the ionization energy differs from reports in literature. \cite{Sato1981,Nielsen1978} Same DBTTF features are weakly observed for the DBTTF-rich film (not shown). However, two new features with onsets at 1.15 eV and 2.80 eV are much more prominent, which are present in the mixed films. This fact points towards the formation of new states that are present only when the two molecular orbitals hybridize and form a DAC. Thus, these features are attributed to HOMO and HOMO$-1$ of the DAC. The spectrum of the HATCN-rich film has strong resemblance with the pristine HATCN film with the addition of weak DAC features. Also, the HATCN HOMO onset is present at the same energy as in the pristine HATCN film. These data are in agreement with phase separation in the non-equimolar films between the DAC and the excess material (DBTTF or HATCN), observed by X-ray scattering and optical spectroscopy results. Taking the HOMO onset of the DAC at 1.15 eV and the optical gap of 1.16 eV (transition energy of the first Taus-Lorentz oscillator), the LUMO of the DAC is located at the value of exciton binding energy above Fermi level. Exciton binding energy are give in the range of several 100\,meV. \cite{Krause2009,Hill2000} This low energy distance between Fermi energy and LUMO position is in accordance with LUMO pinning of the DAC as concluded from the SECO increase mentioned before.

The resulting energies of HOMO onset determined from photoelectron spectroscopy measurements and estimated LUMO values are summarized in Figure~\ref{fig:UPS_Ana}c. For simplification the same pinning level of 0.45 eV for DBTTFs HOMO is used also for the LUMO pinning of DAC and HATCN, as the LUMO pinning levels can only be estimated here. This results in a much lower electron than hole injection barrier for the DAC as well as an "n-type" positioning of $E_F$ within the energy gap of the complex, which will become relevant for electrical transport later.

\subsubsection{Degree of charge transfer}

Clear evidence for the formation of an electron donor-acceptor complex in the mixed\linebreak DBTTF:HATCN films has been obtained from optical absorption and UPS measurements. Here, we analyze the degree of charge transfer from the energy position of the C$\equiv$N vibrational mode as well as the composition of the molecule-specific core electron levels measured by X-ray photoelectron spectroscopy. \cite{Chap81,Grobman1974,Grobman1976}


\begin{figure}[b!]
\begin{center}
\begin{tabular}{c}
\includegraphics[width=\textwidth]{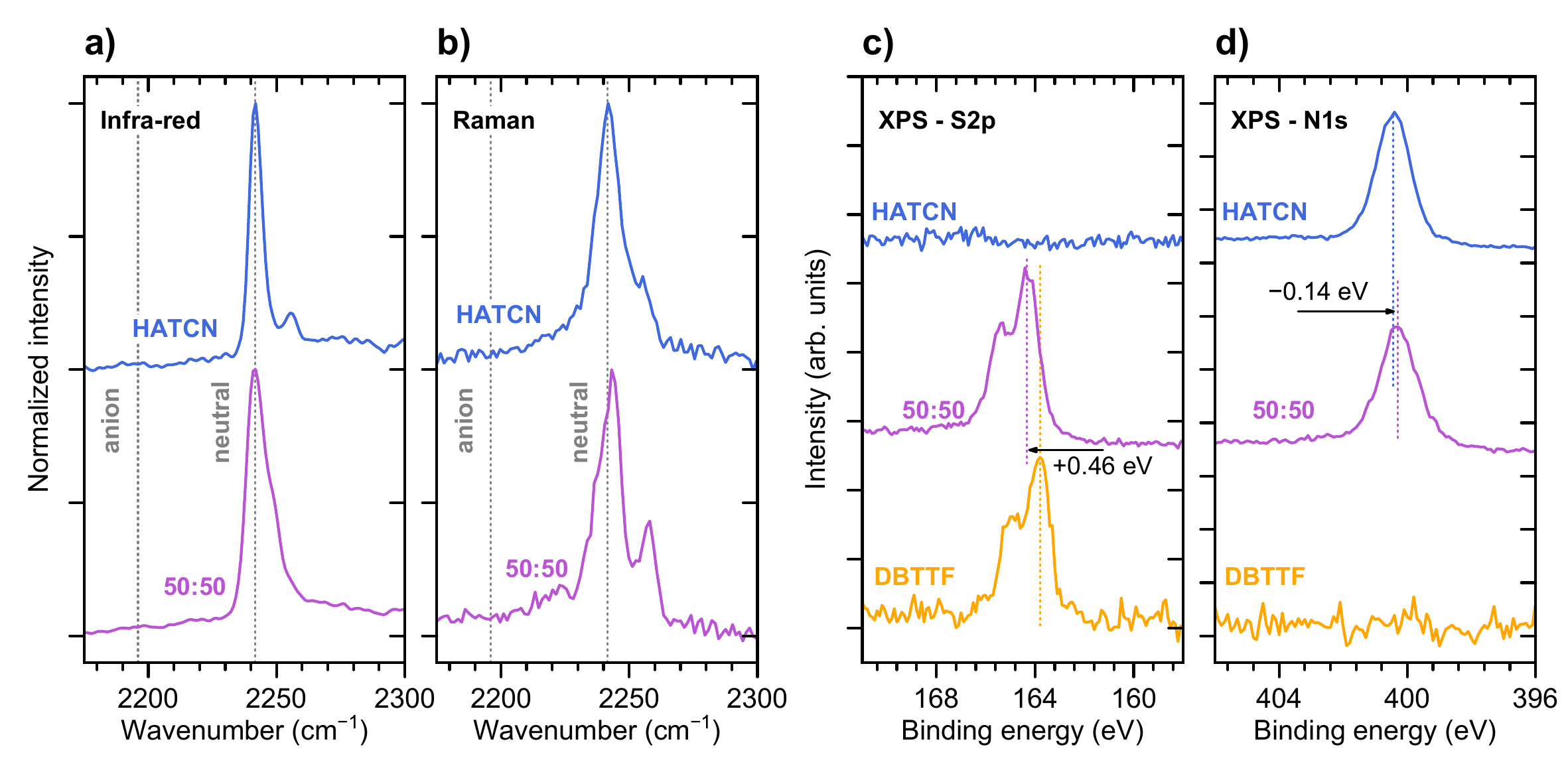}
\\
\end{tabular}
\end{center}
\caption 
{ \label{fig:FTIR}
(a) FTIR (in transmission) and (b) Raman spectra of a HATCN and a DBTTF:HATCN equimolar mixed film in the range of the C$\equiv$N vibration. The mixing fractions are given in mol\%. The mode in the neutral (2241~cm$^{-1}$) and in anionic (2196~cm$^{-1}$) state of the HATCN molecule is taken from literature.\cite{Kona+20CEJ} The X-ray photoelectron spectra (XPS) of (c) S 2p and (d) N 1s peaks for DBTTF, HATCN, and the equimolar mixed thin film. The arrows indicate the peak shifts from the respective peak in the pristine film to the equimolar mixed film.} 
\end{figure} 

The Fourier-transform infrared spectroscopy (FTIR) and Raman spectra of a pristine HATCN film and an equimolar DBTTF:HATCN mixed film are shown in the Figure~\ref{fig:FTIR}a and b, respectively. The neutral HATCN shows an intense Raman line corresponding to the vibrational mode of the C$\equiv$N group at 2241 cm$^{-1}$.\cite{Kona+20CEJ} This line is barely shifted for the DBTTF:HATCN mixture. Also, no significantly shifted new feature is detectable. However, the relative intensity and position of the high energy shoulder are altered in the mixed film compared to the pristine HATCN film. This shows a clear difference in environment of the HATCN molecules in these two films. Given a reported shift of the vibrational mode by 45 cm$^{-1}$  when going from neutral HATCN to anionic HATCN molecules,\cite{Kona+20CEJ} the degree of CT in the DBTTF:HATCN DAC is therefore negligible if a linear relationship between frequency shift and degree of charge transfer is assumed. \cite{Chap81}

The chemical composition of the sample surface was characterized using X-ray photoelectron spectroscopy (XPS). Here the sulfur 2p and the nitrogen 1s orbitals are fingerprints for DBTTF and HATCN, respectively. Figure \ref{fig:FTIR}c-d show the core level energy region of the S 2p and N 1s peaks. Going from pristine films to the equimolar mixed film, the S 2p peak shifts to higher energies and the N 1s peak to lower energies by $+0.46$ eV and $-0.14$ eV, respectively. This is in accordance with the shifts of the DBTTF HOMO and HATCN LUMO levels in forming the new HOMO and LUMO levels of the complex. A detailed deconvolution of the XPS data is given in the SI. In short, the S 2p peak consists of the 2p$_{3/2}$ and the 2p$_{1/2}$ features, while the N 1s peak comprises two features of different bonding environments of nitrogen in HATCN, viz. the different N atoms in the ring and in the cyano groups. Furthermore, both HATCN spectra show a second component shifted to lower binding energies indicating the anionic HATCN species. \cite{Christodoulou2014} As there is no equivalent cationic species found in S 2p spectra of DBTTF, these features are related to charge transfer between HATCN molecules and the substrate, which leads to the above mentioned Fermi level pinning. This effect is much more pronounced for HATCN than for DBTTF films as shown by the larger SECO shift for HATCN than for DBTTF. Consequently, the intermolecular CT between DBTTF and HATCN is negligible, in agreement with the results from vibrational spectroscopy.
Note that, in general, the degree of CT in donor-acceptor cocrystals varies between different polymorphs, and zero degree of CT was also observed for other systems, like DBTTF:TCNQ and DBTTF:TCNNQ. \cite{beye+19cm,Goet+17AEM}

\subsubsection{Electrical conductivity}

The electrical conductivity of organic semiconductors is often governed by hopping transport, wherein the conductivity exhibits a strong temperature dependence, displaying thermally activated behavior as described by the Arrhenius model (Equation \ref{eq_Arr1}). This behavior is characterized by a material-specific activation energy $E_{a}$ and a prefactor $\sigma_{0}$, which represents the conductivity limit at an idealized infinite temperature.
\begin{equation}\label{eq_Arr1}
\sigma(T) = \sigma_{0} \cdot \exp\left(-\dfrac{E_{a}}{k_{B}T}\right)
\end{equation}
Note that this equation can be used in a two-fold manner: First, to extract the activation energy of a given sample from its temperature-dependent electrical conductivity, yielding values of $E_a$ and $\sigma_0$ for each different composition. And, secondly, to compile a systematic relation between the conductivities of different samples at a given temperature, e.g. $\sigma(300\,\mathrm{K})$, and their activation energies, as will be discussed below.

Improving electrical conductivity necessitates either a reduction in the activation energy $E_a$ or an enhancement of $\sigma_{0}$. Recent findings indicate that a broad range of doped organic semiconductors with integer charge transfer adhere to this model at room temperature ($T$=300 K), with a common $\sigma_{0}$ value around 15 S/cm, while activation energies range from 40 to 400 meV.\cite{Schwarze19}

\begin{figure}[tb]
\begin{center}
\begin{tabular}{c}
\includegraphics[width=15cm]{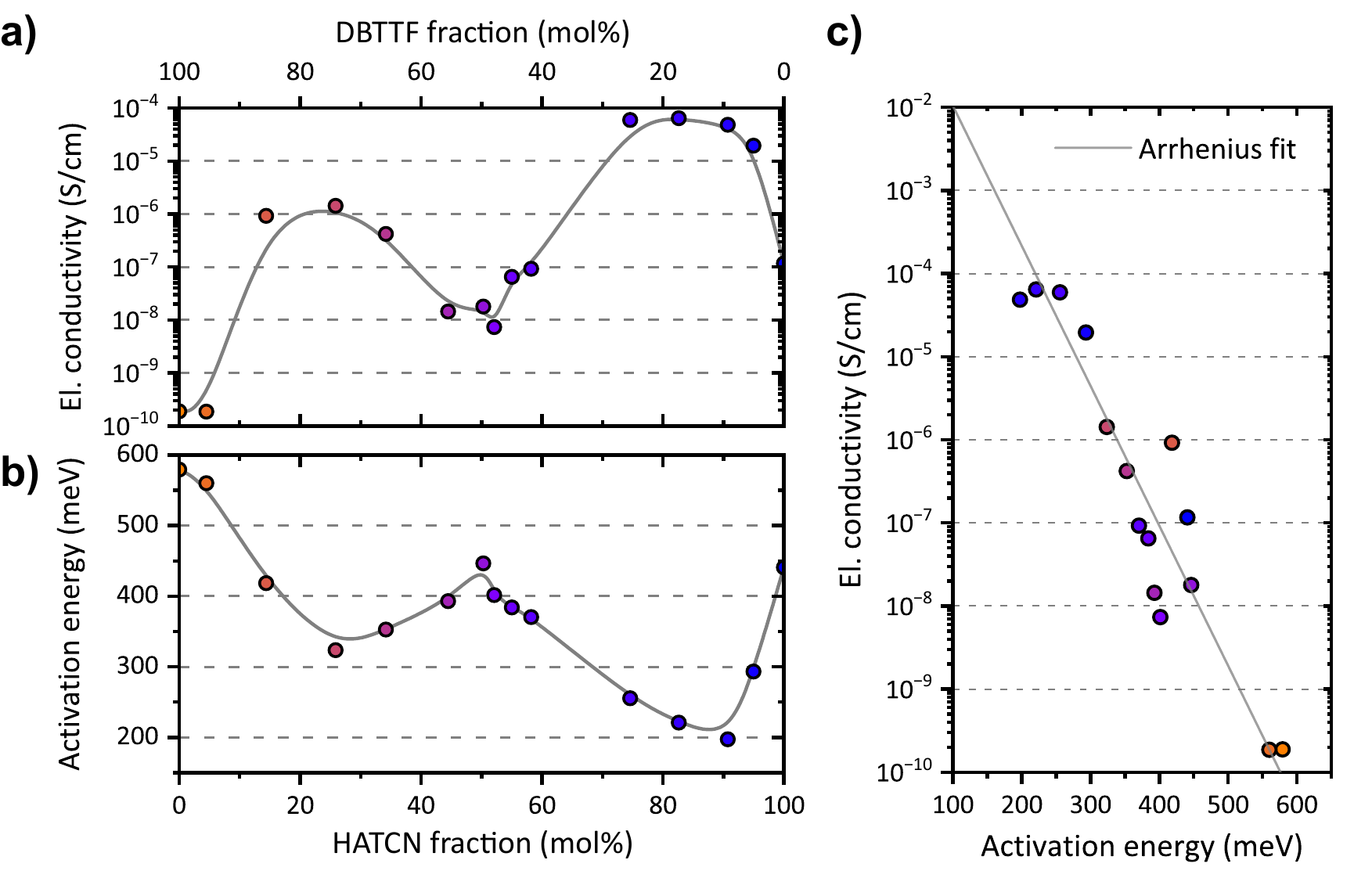}
\\
\end{tabular}
\end{center}
\caption 
{ \label{fig:Conductivity}
(a) Electrical conductivity (at room temperature) and (b) thermal activation energy (from temperature-dependent conductivity measurements) across the entire range of DBTTF:HATCN concentrations. The data exhibit an inverse relationship between electrical conductivity and activation energy, consistent with the Arrhenius model. (c) The room-temperature conductivity data are fitted using the Arrhenius equation (\ref{eq_Arr1}), yielding a prefactor $\sigma_0=1.2\,\mathrm{S/cm}$.}
\end{figure} 

To investigate the behavior of DBTTF:HATCN within this context, we prepared 15 distinct mixtures of the two materials, including pristine films of each species. Figure \ref{fig:Conductivity} illustrates the room-temperature conductivities and thermal activation energies across the entire mixing range. Detailed Arrhenius fit parameters for all samples are provided in Table~S1 in the Supporting Information.

Notably, the electrical conductivity of DBTTF:HATCN mixtures at 300 K (Figure \ref{fig:Conductivity}a) reveals two maxima at approximately 25 and 80\,mol\% HATCN. Correspondingly, there are two minima in activation energy (Figure \ref{fig:Conductivity}b), with the lowest value of about 200 meV observed at 80-90\,mol\% HATCN. Interestingly, the equimolar DBTTF:HATCN mixture exhibits a local conductivity minimum and an activation energy maximum. It should be noted that pristine DBTTF (and the 95\,mol\% DBTTF films) have extremely low conductivity, attributed to non-continuous film formation, leading to island growth as evidenced by AFM imaging.

Figure \ref{fig:Conductivity}c presents room temperature conductivity plotted against thermal activation energy for all samples. Despite some scattering, the data appear to be consistent to the Arrhenius equation, as indicated by the fitted linear trend. This means that, similar to the case of ICT\cite{Schwarze19}, there is a correlation between room temperature conductivity and activation energy in these samples. However, the temperature independent prefactor, $\sigma_{0}$, obtained from the fit is somewhat lower (1.2 S/cm) than for ICT. This could indicate a lower doping efficiency in DBTTF:HATCN.

\begin{figure}[tb]
\begin{center}
\begin{tabular}{c}
\includegraphics[width=0.75\linewidth]{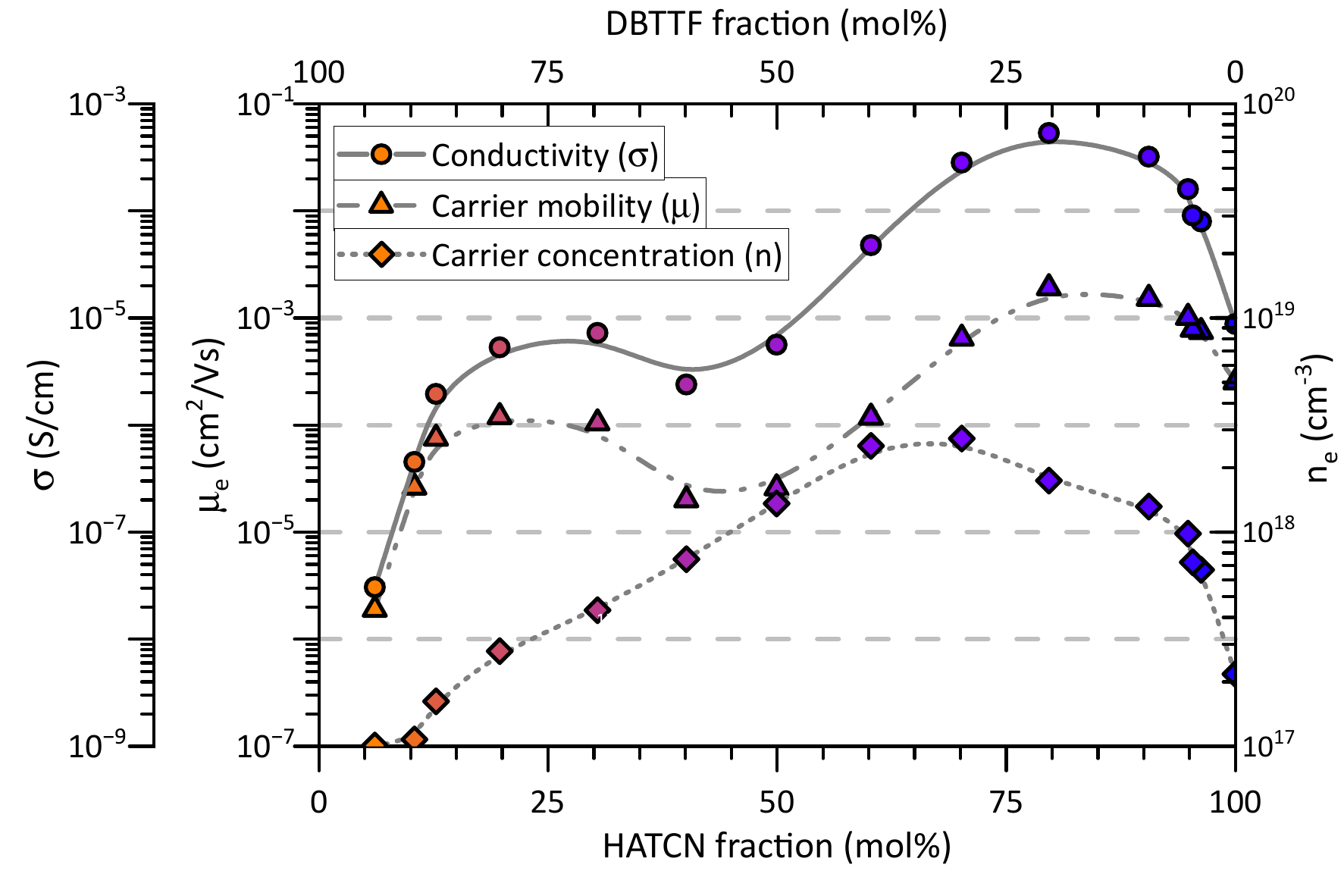}
\\
\end{tabular}
\end{center}
\caption 
{ \label{fig:Mobility}
Charge carrier mobility ($\mu$) and carrier concentration ($n$) in DBTTF:HATCN thin films were extracted from field-effect transistor measurements and from Mott-Schottky analysis on MIS diodes. In both experimental approaches only negative charge carriers could be detected. The product of both quantities yields the electrical conductivity ($\sigma$).} 
\end{figure} 

In addition to direct conductivity measurements, we also determined charge carrier mobility ($\mu$) from organic field-effect transistor (OFET) characteristics as well as carrier density ($n$) and carrier type from Mott-Schottky analysis on metal-insulator-semiconductor (MIS) diodes. The raw data is presented in Figure~S6 in the Supplementary Information, while the extracted mobility and carrier density values are displayed in Figure \ref{fig:Mobility}. Due to morphological constraints, pristine DBTTF films were not measurable. Remarkably, all samples display n-type behavior irrespective of mixing fraction, with carrier concentration peaking at approximately 60--70\,mol\% HATCN content.
At the same time, carrier mobility in DBTTF:HATCN thin films exhibits a double-maximum at around 20 and 80\,mol\% of HATCN, with a local minimum at the equimolar mixing ratio. As per the relationship $\sigma = q n_e\mu_e$, where $q$ is the elementary charge, we calculated the resulting electrical conductivity. This calculated conductivity mirrors the experimental values, albeit with less variability and a less pronounced minimum in the range of equimolar mixtures. It is important to note that OFET and MIS diode measurements assess orthogonal sample directions (in-plane and out-of-plane, respectively), which could account for discrepancies with direct in-plane conductivity measurements. Additionally, the top-contact configuration employed in the MIS diode measurements of charge carrier concentration and mobility may introduce differences in contact resistance and anisotropic mobilities compared to the bottom-contact setup used in direct electrical conductivity measurements. Nevertheless, the agreement with direct conductivity measurements shown in Figure \ref{fig:Conductivity}a is very good.

\section{Discussion}

The classification of organic semiconducting materials as donors or acceptors relies mostly on their relative energy levels. Typically, doping of one species by the other is performed by adding small amounts (1--10\%) of a strong acceptor to a donor matrix to achieve p-type doping, or vice versa for n-type doping. Here we vary the mixing ratio of such a donor-acceptor pair over the whole composition range from the pristine donor (DBTTF) to the pristine acceptor (HATCN) and study its phase behavior as well as the implications on charge transport of such mixtures.

While we would -- a priori -- expect a (more or less) symmetric behavior between the donor-rich (p-doped) side and the acceptor-rich (n-doped) side, a strong asymmetry is found in the studied case of DBTTF:HATCN. Specifically, pristine DBTTF shows very low (or almost negligible) electrical conductivity, while high conductivities are found on the HATCN-rich side (Figure~\ref{fig:Conductivity}a). Moreover, except for the pristine DBTTF all compositions show n-type conduction. However, the increase of conductivity from the donor-rich to the acceptor-rich side is non-monotonic but shows a pronounced conductivity minimum at around equimolar composition.

\begin{figure}[htbp]
\begin{center}
\includegraphics[width=\textwidth]{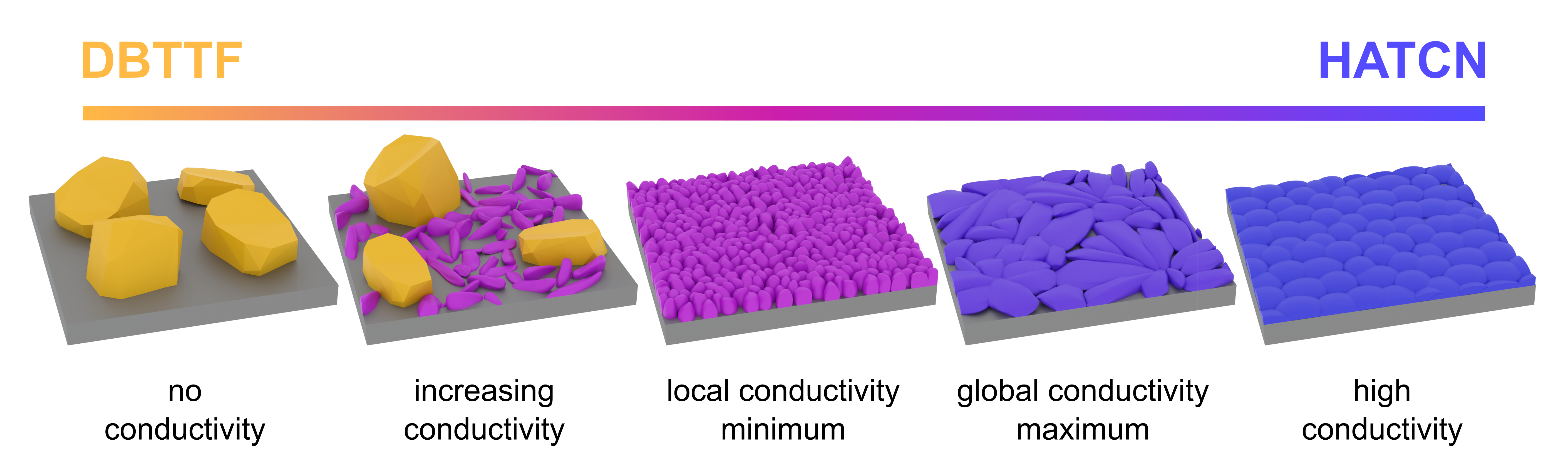}
\end{center}
\caption{Sketch of the phase behavior of DBTTF:HATCN mixtures and the implications for charge transport along the composition gradient from pristine DBTTF (left) to pristine HATCN (right). DBTTF crystallites are shown in dark yellow, the DAC in purple and HATCN in dark blue. Further details are given in the text.
\label{fig:phase-behavior}}
\end{figure}

To account for this unexpected behavior, we have performed a comprehensive series of experiments combining film growth and device studies to elucidate the phase behavior of the mixtures and relate it to the observed electrical transport. This is sketched schematically in Figure \ref{fig:phase-behavior}.
In particular, film structure and morphology studies indicate that pristine DBTTF forms large almost non-percolating crystallites, which persist over the entire DBTTF-rich side of the composition range and lead to considerable film roughness, whereas the HATCN-rich side exhibits very smooth films without pronounced signs of phase separation (Figure~\ref{fig:AFM}). Most importantly, however, apart from the known pristine DBTTF and HATCN crystal structures, we find a distinctly different new crystal structure around the equimolar mixing ratio of both components, which we assign to a donor-acceptor complex (Figure~\ref{fig:GIWAXS}). This is corroborated by optical absorption spectra that unequivocally reveal the appearance of a new absorption feature at about 1.3-1.4\,eV below the absorption edges of the two pristine materials, which is a clear fingerprint of a charge-transfer state formed by electronic interaction between the DBTTF donor and the HATCN acceptor (Figure~\ref{fig:UV-Vis}).

Photoelectron spectroscopy further manifests this view (Figure~\ref{fig:UPS_Ana}). The DAC exhibits two new, clearly separated occupied states, which we assign to HOMO and HOMO-1 of the complex. Their formation can be envisioned by hybridization between the respective occupied states of the donor DBTTF and the LUMO of HATCN. Importantly as for the latter, the LUMO of the DAC seems to be pinned at the same relative energy of -0.45\,eV above the Fermi level of the ITO substrate, which may be seen as the root origin of the observed n-type conduction in all the mixtures, since the barrier for hole injection into the HOMO of the complex is way too high ($>1.1$\,eV).

Interestingly, however, infrared/Raman as well as X-ray photoelectron spectroscopy do not indicate a substantial degree of ground state charge transfer between DBTTF and HATCN in the formed DAC (Figure~\ref{fig:FTIR}). \cite{Rußegger2022} Nevertheless, we observe an increase of carrier density (Figure~\ref{fig:Mobility}) from Schottky-Mott analysis on MIS diodes that is clearly correlated with the existence of the new crystal structure of the DAC and its optical signature in UV/Vis absorption. Note, the DAC cocrystals act as dopants here.\cite{Henry15}  The carrier density peaks at about 60--70\% HATCN, similar to the other two signatures, and can be clearly assigned to the formation of the DAC. Thus, even without noticeable CT between donor and acceptor, the complex promotes the generation of mobile charge carriers, which is the prerequisite for an enhancement of electrical conductivity in the mixture.

What is, nonetheless, somewhat unexpected, is that the electrical conductivity does not just follow the carrier density but exhibits a local minimum around the equimolar mixing ratio of DBTTF and HATCN, which is also reflected in a minimum of charge carrier mobility in this range (Figure~\ref{fig:Mobility}). This "anomaly" seems to be related to film morphology, as evidenced by our AFM studies. Starting from pristine DBTTF with its pronounced island growth and very rough films (Figure~\ref{fig:AFM}h), the increasing amount of the DAC (growing as rather smooth "filler" between the DBTTF islands, Figure~\ref{fig:AFM}f-g) improves all electrical parameters, including carrier density and mobility, by forming a percolating crystalline network in between the DBTTF islands. However, as soon as the equimolar ratio is reached with the new crystalline phase of DAC being the dominant species, the grain size is shrinking and causes a very fine-grained film morphology with lots of grain boundaries (Figure~\ref{fig:AFM}e). We ascribe this change in film morphology, which is coupled to the prevalence of the DAC, to the main cause for the reduction in charge carrier mobility and electrical conductivity in equimolar mixtures. 
The global maximum in electrical conduction is reached at about 70--90\% HATCN content in the mixture because, in this range, film morphology exhibits very crystalline needle-like structures (Figure~\ref{fig:AFM}c), which are favorable for charge transport. At the same time, the signature of DAC crystallites rapidly vanishes so that the remaining DACs are probably molecularly dispersed in the HATCN matrix and the DAC cocrystals, as mentioned before, act as n-dopants.

Overall, electrical conductivity and its thermal activation show a similar behavior as it was found for integer CT doping in a wide range of organic semiconductors (Figure~\ref{fig:Conductivity}c). It follows an Arrhenius law with a "universal" prefactor that that is roughly by a factor of 10 smaller compared to the ICT case. This can be  considered as an indicator for a lower doping efficiency in complex doping. \cite{beye+19cm,opitz2020ordered}

\section{Conclusion}

We have studied the structural phase and mixing behavior of the organic molecular donor DBTTF and the acceptor HATCN. Using a dedicated co-deposition technique, we were able to fabricate samples with a seamless composition gradient on a single substrate, which are used for high-throughput analysis of film properties. By correlating the obtained structural, morphological and optical data with electrical and electronic device characteristics, we were able to identify the formation of a DAC with a distinctly different crystal structure and characteristic electronic signatures in the subgap region of the pristine materials. Electrical transport is affected in two ways by DAC formation: On the one hand, DACs increase carrier density and their LUMO-level pinning leads to n-type conduction over almost the entire composition range. On the other hand, the prevalence of the new DAC phase around equimolar composition is characterized by a fine-grained morphology with the abundance of grain boundaries, which in turn reduces the macroscopic charge carrier mobility and leads to a pronounced minimum in electrical conductivity at this composition.

Altogether, our comprehensive studies of the phase behavior of this DAC forming mixture provide detailed insights into their electrical transport and its correlation to morphology, which is the prerequisite for developing new materials with tailored electrical and optical properties, e.g. infrared photodetectors.


\begin{suppinfo}
Experimental details, additional information on growth study, absorption, film morphology for devices, core level structure and electrical transport.
\end{suppinfo}

\section{Notes}
The authors declare no competing financial interest.

\section{Data availability statement}
Data produced at the ID10-SURF beamline at ESRF (Grenoble, France) are available at \url{doi.org/10.15151/ESRF-ES-1940867632}.

\section{Acknowledgments}
Deutsche Forschungsgemeinschaft (project numbers 239543752 and 530008779) supported this work financially. This study is partially funded by the Federal Ministry of Education and Research (BMBF) and the Baden-Württemberg Ministry of Science as part of the Excellence Strategy of the German Federal and State Governments.
We acknowledge the European Synchrotron Radiation Facility (ESRF) for provision of synchrotron radiation facilities under proposal number SC-5641."
L.F. and M.S. gratefully acknowledge financial support from the German Environmental Foundation (Deutsche Bundesstiftung Umwelt, DBU).
J.P. acknowledges financial support from the Bavarian State Ministry for Science and the Arts within the collaborative research network “Solar Technologies go Hybrid (SolTech)”. L. S.-M. appreciates support from the Würzburg-Dresden Cluster of Excellence on Complexity and Topology in Quantum Matter, ct.qmat (EXC 2147).

\bibliography{report}

\end{spacing}

\begin{tocentry}
		\includegraphics[width=\textwidth]{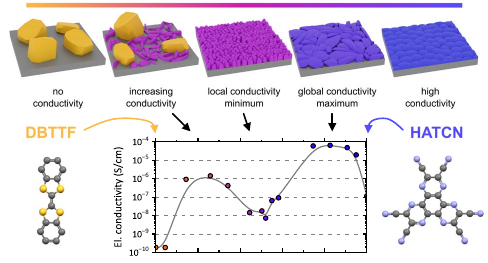}
\end{tocentry}
\end{document}